\PassOptionsToPackage{unicode}{hyperref}
\PassOptionsToPackage{hyphens}{url}
\documentclass[
]{article}
\usepackage{xcolor}
\usepackage{amsmath,amssymb}
\usepackage{setspace}
\usepackage{iftex}
\usepackage{graphicx}
\ifPDFTeX
  \usepackage[T1]{fontenc}
  \usepackage[utf8]{inputenc}
  \usepackage{textcomp} 
\else 
  \usepackage{unicode-math} 
  \defaultfontfeatures{Scale=MatchLowercase}
  \defaultfontfeatures[\rmfamily]{Ligatures=TeX,Scale=1}
\fi
\usepackage{lmodern}
\ifPDFTeX\else
\fi
\IfFileExists{upquote.sty}{\usepackage{upquote}}{}
\IfFileExists{microtype.sty}{
  \usepackage[]{microtype}
  \UseMicrotypeSet[protrusion]{basicmath} 
}{}
\makeatletter
\@ifundefined{KOMAClassName}{
  \IfFileExists{parskip.sty}{%
    \usepackage{parskip}
  }{
    \setlength{\parindent}{0pt}
    \setlength{\parskip}{6pt plus 2pt minus 1pt}}
}{
  \KOMAoptions{parskip=half}}
\makeatother
\usepackage{longtable,booktabs,array}
\usepackage{multirow}
\usepackage{calc} 
\usepackage{etoolbox}
\makeatletter
\patchcmd\longtable{\par}{\if@noskipsec\mbox{}\fi\par}{}{}
\makeatother
\IfFileExists{footnotehyper.sty}{\usepackage{footnotehyper}}{\usepackage{footnote}}
\makesavenoteenv{longtable}
\usepackage{graphicx}
\makeatletter
\newsavebox\pandoc@box
\newcommand*\pandocbounded[1]{
  \sbox\pandoc@box{#1}%
  \Gscale@div\@tempa{\textheight}{\dimexpr\ht\pandoc@box+\dp\pandoc@box\relax}%
  \Gscale@div\@tempb{\linewidth}{\wd\pandoc@box}%
  \ifdim\@tempb\p@<\@tempa\p@\let\@tempa\@tempb\fi
  \ifdim\@tempa\p@<\p@\scalebox{\@tempa}{\usebox\pandoc@box}%
  \else\usebox{\pandoc@box}%
  \fi%
}
\def\fps@figure{htbp}
\makeatother
\usepackage{bookmark}
\IfFileExists{xurl.sty}{\usepackage{xurl}}{} 
\hypersetup{
  hidelinks,
  pdfcreator={LaTeX via pandoc}}

\author{}
\date{}

\begin{document}
\setstretch{1.2}

\title{\Huge Modeling Brownian Motion As A Timelapse Of The Physical, Persistent, Trajectory}
\maketitle
\begin{center}
\emph{\large Ludovico Cademartiri}
\end{center}
\emph{\small Department of Chemistry, Life Sciences and
Environmental Sustainability, University of Parma, Parco Area delle
Scienze 17 A, Parma, Italy}
\begin{center}
\emph{\small ludovico.cademartiri@unipr.it}
\end{center}

\subsection{Abstract}\label{abstract}

While it is very common to model diffusion as a random walk by assuming
memorylessness of the trajectory and diffusive step lengths, these
assumptions can lead to significant errors. This paper describes the
extent to which a physical trajectory of a Brownian particle can be
described by a random flight.

Analysis of simple timelapses of physical trajectories (calculated over
collisional timescales using a velocity autocorrelation function that
captures the hydrodynamic and acoustic effects induced by the solvent)
gave us two observations: (i) subsampled trajectories become genuinely
memoryless only when the time step is \textasciitilde200 times larger
than the relaxation time, and (ii) the distribution of the step lengths
has variances that remain significantly smaller than 2Dt (usually by a
factor of \textasciitilde2).

This last fact is due to the combination of two effects: diffusional MSD
is mathematically superballistic at short timescales and subsampled
trajectories are moving averages of the underlying physical trajectory.
In other words, the MSD of the physical trajectory at long time
intervals asymptotically approaches 2Dt, but the MSD of the subsampled
steps does not, even when the associated time interval is several
hundred times larger than the relaxation time. I discuss how to best
account for this effect in computational approaches.

\subsection{Introduction}\label{introduction}

Ensembles of objects in liquids play a major role in questions of great
practical and fundamental importance: how does the nucleation of
crystals occur? how do proteins fold and interact with ligands,
surfaces, and each other? how does diffusion work in crowded
environments? A \emph{trait d'union} of these questions is dynamics,
i.e., how do those objects move under mutual influence and give rise to
collective effects?

One of the challenges with the study of such ensembles is that, on one
hand, they are often too numerous and composed of objects too large to
be conveniently described at an atomic
scale\hyperref[_ENREF_1]{\textsuperscript{1-3}}; on the other hand, they
are often composed of objects too small to be accurately described by
most Brownian dynamics
approaches\hyperref[_ENREF_4]{\textsuperscript{4}}. One strategy to
approach this problem is to make computation at the atomic scale more
efficient\hyperref[_ENREF_5]{\textsuperscript{5}} or to find better
approximations of mean fields\hyperref[_ENREF_6]{\textsuperscript{6}}.
Another strategy, which I espouse here, is to make Brownian descriptions
more accurate and practical with smaller
objects\hyperref[_ENREF_7]{\textsuperscript{7}}.

As part of my interest in understanding the diffusional dynamics of
ensembles, I looked for approaches to Brownian dynamics simulations of
large molecules/nanoparticles in a way that captures the correlations in
the system better than conventional memoryless random walk approaches:
this paper is the account of that work.

In summary (cf. Figure 1), I first calculate a ``physical trajectory'',
defined here as the trajectory of a Brownian particle calculated at
collisional timescales and that considers most correlation effects. Then
I take a ``timelapse'' of this trajectory at time intervals that are
small multiples of the relaxation times. The result is a coarser-grained
trajectory, a flight whose autocorrelation functions are ``steep''
enough to easily provide the autoregression coefficients. These
coefficients are what then allows to easily generate ex novo
trajectories that encode the correlations of the real, physical
trajectory.

\begin{figure}
    \centering
    \includegraphics[width=1\linewidth]{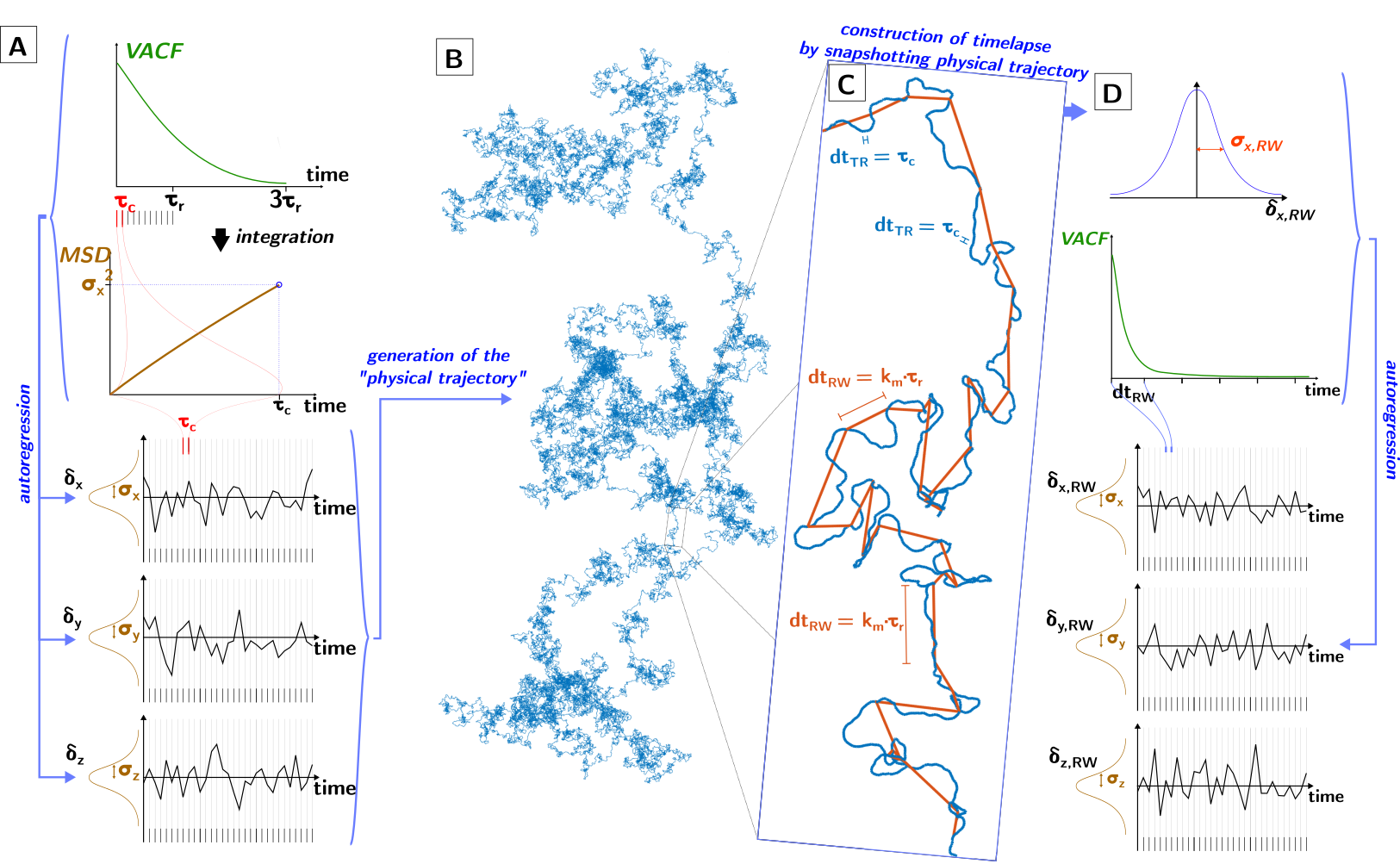}
    \caption{\textbf{Timelapse of the physical trajectory. (A-B)}
\emph{Generation of the physical trajectory.} From the top down, VACF is
calculated up to 3$\tau$\textsubscript{\!r} on $\tau$\textsubscript{\!c} intervals.
The MSD at $\tau$\textsubscript{\!c} is calculated from a finely integrated
VACF from 0 to $\tau$\textsubscript{\!c}. The autoregression (AR) coefficients
of the long-term VACF, together with the MSD at $\tau$\textsubscript{\!c} are
used to generate autocorrelated time series of the cartesian components
of the displacement $\delta$\textsubscript{x}, $\delta$\textsubscript{y},
$\delta$\textsubscript{z} at $\tau$\textsubscript{\!c} intervals. \textbf{(B-C)}
\emph{Construction of the timelapse.} As the physical trajectory is
generated (blue line), snapshots are taken at time intervals
\emph{$\tau$\textsubscript{\!K}} integer multiples of
\emph{$\tau$\textsubscript{\!r}}. The assembled timelapses constitute a
segmented path (orange line). \textbf{(C-D)} \emph{Analysis of
timelapse.} The timelapse is analyzed as a persistent random flight,
extracting the MSD and VACF. From those, the autoregression coefficients
are finally obtained which allow for the rapid generation ex novo of
persistent random flights that accurately match the physical trajectory.}
    \label{fig:one}
\end{figure}

Particles in liquid solutions display a spontaneous net displacement
over time. Lucretius already intuited in 58 BC that this motion was
evidence of the existence of
atoms\hyperref[_ENREF_8]{\textsuperscript{8}}, but it took nearly 2000
years to develop a sound theory of
it\hyperref[_ENREF_9]{\textsuperscript{9-11}}. This motion is subject to
a set of conditions: (i) potential gradients are finite, i.e., forces
are finite, i.e., accelerations are finite, i.e., the trajectory must be
continuous; (ii) the mass of each particle in the liquid is finite and
typically larger than that of the individual liquid molecules
surrounding it, i.e., motion is to some degree inertial; (iii) kinetic
energy is Maxwell-Boltzmann-distributed, consistently with the laws of
thermodynamics and the equipartition theorem; (iv) the structure of
liquid solutions is dynamic, disordered in space but homogeneous and
isotropic over time, and everything is effectively ``in contact'' with
its neighbors; (v) energy barrier confining a particle in a position of
the liquid is finite; (vi) thermal motion is uncorrelated. Based on
these facts and assumptions, a particle suspended in a liquid does not
oscillate in place but rather diffuses, i.e., its root mean square (RMS)
displacement grows with time.

In what follows I only consider spherical objects in the absence of
advective flows. Non-sphericity and shear introduce significant
complications in the description of motions, collisions, and
viscosity\hyperref[_ENREF_12]{\textsuperscript{12-15}}. For convenience
I also describe the distinct objects that make up the liquid as
``solvent molecules'' (even though there is no requirement for them to
be molecules), and the object being moved by collisions with the solvent
molecules as the ``particle'' (even though there is no requirement for
it to be a solid phase and not a molecule).

The motion of particles in solution is characterized by two distinct
timescales\hyperref[_ENREF_16]{\textsuperscript{16-19}}.

\textbf{Collision time.} The ``collisional'' timescale is defined here
as the mean time that separates two solvent/particle ``collisions''. As
there are no real collisions in liquids (nearest neighbors are always
``in contact''), by ``collision'' I intend here any transfer of momentum
between two neighboring particles caused by orbital overlap. I estimate
the mean time between solvent/particle collisions by using a simple
logic.

If I assume a weakly interacting liquid with a structure close to random
close packing the average number of solvent molecules in contact with
the particle is $\sim 0.886 \cdot ( \tfrac{r_p}{r_s} )^{2} $,
where \emph{r\textsubscript{p}} is the radius of the particle,
\emph{r\textsubscript{s}} is the radius of a solvent molecule, and 0.886
is the packing fraction for a random close-packed arrangement of circles
on a surface\hyperref[_ENREF_20]{\textsuperscript{20}}.

Any displacement that could trigger a transfer of momentum by
``contact'' must cause a sufficient overlap between atomic orbitals and
therefore should be in the order of an atomic radius, i.e.
\textasciitilde1 Å. Therefore, I am assuming the absence of ions and
electrolytes that can interact over large distances. For each solvent
molecule on the surface of the particle, the RMS time it takes to move
by 1 Å is:

\[
\frac{1 \cdot 10^{-10}}{v_{\text{RMS}}} = \dfrac{1 \cdot 10^{-10}}{\sqrt{\dfrac{3 k_B T}{m_s}}}
\]

Where v\textsubscript{RMS} is the RMS velocity of the solvent molecules,
k\textsubscript{B} is Boltzmann constant, T is the temperature and
m\textsubscript{s} is the mass of a solvent molecule.

Considering that during this time a solvent molecule can move towards
the particle or away from it, and that the particle can also move
causing a collision with ½ of the particles on its surface, the
frequency f\textsubscript{c} of collisions between the solvent molecules
on the surface and the particle can be estimated as

\[
f_c \approx \frac{\sqrt{3 k_B T}}{2.26 \cdot 10^{-10}} \left( \frac{r_p}{r_s} \right)^2 \left( \frac{1}{\sqrt{m_s}} + \frac{1}{\sqrt{m_p}} \right) \tag{2}
\]

Which means that the mean time $\tau$\textsubscript{\!c} between two
consecutive solvent/particle collisions can be estimated as its inverse

\[
\tau_c \approx \frac{2.26 \cdot 10^{-10}}{\sqrt{3 k_B T}} \left( \dfrac{r_s}{r_p} \right)^2 \left( \dfrac{\sqrt{m_s m_p}}{\sqrt{m_s} + \sqrt{m_p}} \right) \tag{3}
\]

This timescale is most important: if one wants to describe the effect of
these solvent/particle collisions as an effective viscous force, one
needs to consider timescales $\gg$ $\tau$\textsubscript{\!c} so that the number of
collisions occurring at each step is large enough to give a stable mean
effect.

The largest contributor to $\tau$\textsubscript{\!c} is the size of the
particle, with the physical properties of the solvent and the
temperature being minor effects (cf. Figure S1).

\textbf{Relaxation time.} While $\tau$\textsubscript{\!c} tells us at what
timescales we can make a continuum approximation of the solvent's
viscous drag, the effect that each solvent/particle collision has on the
dissipation of the momentum of the particle must depend on its inertia
and, therefore, on its density.

In the first order approximation in which the viscous force $\chi$ acting on
the particle is proportional to its velocity, the decay of particle's
momentum occurs exponentially with a decay constant $\tau$\textsubscript{\!r},
called the ``relaxation
time''\hyperref[_ENREF_19]{\textsuperscript{19}},
\[
\tau_r = \frac{m_p}{\chi} \; ; \quad \text{for spheres: } \tau_r = \frac{2}{9} \frac{\rho_p}{\eta} r_p^2 \tag{4}
\]

where $\rho$\textsubscript{p} is the particle density and $\eta$ is the dynamic
viscosity of the solvent. This relaxation time quantifies the timescale
over which the momentum of the particle is dissipated/decorrelated (both
in direction and magnitude) by the collisions with the solvent
molecules.

These two timescales have opposite dependencies on the size of the
particle: $\tau$\textsubscript{\!c} goes with
r\textsubscript{p}\textsuperscript{-2} while $\tau$\textsubscript{\!r} goes
with r\textsubscript{p}\textsuperscript{2}. This fact implies that,
depending on the solvent and on the density of the particles, there is a
particle radius below which the approximations I have made so far cease
to make physical sense (cf. Figure S2). Physically speaking, below a
certain particle size (and hence, mass) it takes very few collisions
with the solvent molecules, even one, to significantly dissipate the
momentum of the particle. In these conditions viscosity cannot describe
dissipation, Equation 4 ceases to be valid, and a full collisional
description of the system is required. This critical particle radius
(cf. Figure S2) lies in the nanoscale for the most common solvents and
solid phases.

\textbf{The random flight approximation.} In most computational and
theoretical studies of Brownian motion the motion of the particles is
approximated as a random flight whose steps occur over fixed time
intervals\textsuperscript{\hyperref[_ENREF_21]{21},\hyperref[_ENREF_22]{22}}.
Therefore, the choice of the time interval is crucial.

Beside $\tau$\textsubscript{\!r} and $\tau$\textsubscript{\!c}, there is a third
timescale that is often employed in the modeling of ensembles of
Brownian objects, which is the time $\tau$\textsubscript{\!e} it takes for the
RMS displacement (usually assumed to be diffusive) to equal a radius of
the particle under consideration (i.e., for spheres, $\tau$\textsubscript{\!e}
= r\textsubscript{p}\textsuperscript{2}/6D).

As I show here, while it is correct to assume that, for a Brownian
particle (i.e., one for which $\tau$\textsubscript{\!r} $\gg$ $\tau$\textsubscript{\!c}),
the VACF has fully disappeared over $\tau$\textsubscript{\!e} (i.e., it is
legitimate to use a non-persistent random flight over such timescales),
it is quite incorrect, even at such timescales, to use a gaussian
distribution with 2Dt variance to describe the cartesian components of
the steps. The real step lengths are quite a bit shorter because of the
effect of autocorrelation at short timescales (i.e., ballistic motion is
mathematically subdiffusive at short enough timescales: the diffusive
velocity \(\frac{\sqrt{6D}}{t}\ \)→$\infty$ as t→0).

If one chooses (e.g., to consider effects of shape) to look at finer
timescales of motion (timesteps in the order of small multiples of
$\tau$\textsubscript{\!r}) the autocorrelation must be considered explicitly
and its effect on the trajectory is nontrivial.

\subsection{Computational Design}\label{computational-design}

\textbf{Computational parameters.} Our simulations were conducted
considering a library of relevant solvents (H\textsubscript{2}O, EtOH,
toluene, tetrachloroethylene, and hexadecane) and particle phases
(proteins, amorphous SiO\textsubscript{2}, calcite,
Fe\textsubscript{3}O\textsubscript{4}, CdSe, and Au) and particle radii
\emph{r\textsubscript{p }}between 1.66·10\textsuperscript{-10} m and
1·10\textsuperscript{-8} m in 10 logarithmically distributed steps.

Out of all possible combinations of these parameters, I excluded from
consideration those for which m\textsubscript{s} \textgreater{}
m\textsubscript{p} or $\tau$\textsubscript{\!r} \textless{}
10·$\tau$\textsubscript{\!c}. This last condition ensures that at least 10
solvent/particle collisions are needed to decorrelate the motion of the
particle from 1 to \textasciitilde1/e. Translating this condition in an
inequality between the properties of particle and of the solvent
molecules the condition leads to

\[
\rho_p \cdot r_p^{4} \gtrsim 1.098 \cdot 10^{-8} \frac{\eta r_s^{2}}{v_{\text{RMS},s}} = \Gamma_K \tag{5}
\]

The right side of the inequality is a temperature-dependent property of
the solvent ($\Gamma$\textsubscript{K}, in {[}Kg·m{]}) that can be easily
tabulated for reference (cf. Table I)

\textbf{Table I.}

\begin{table}[ht]
    \centering
    \begin{tabular}{lll}
        \toprule
        \textbf{Solvent} & \textbf{Temperature [°C]} & \textbf{$\boldsymbol{\Gamma_K}$ [10\textsuperscript{-33} kg·m]} \\ 
        \midrule
        Water           & 25                        & 0.419                         \\
                        & 70                        & 0.391                         \\ 
        \midrule
        Toluene         & 25                        & 1.95                          \\ 
        \midrule
        Ethanol         & 25                        & 1.77                          \\ 
        \midrule
        Tetrachloroethylene & 25                    & 3.75                          \\ 
        \midrule
        Hexadecane      & 25                        & 32.5                          \\
                        & 300                       & 23.5                          \\ 
        \bottomrule
    \end{tabular}
    \label{tab:solvent_data}
\end{table}

Using these values it is easy to estimate the smallest particle radius
r\textsubscript{B} that could be reasonably described by a Brownian
simulation (with the possible caveat of having to consider
autocorrelation effects). For a SiO\textsubscript{2} particle in water
r\textsubscript{B} is 0.654 nm, for CdSe in hexadecane at 300 C (close
to synthesis conditions for colloidal quantum dots) r\textsubscript{B}
is 0.693 nm, while for Au particles in water is 0.384 nm. Importantly,
for a protein in water at room temperature (considering a
molecular-weight independent density of 1220 Kg/m\textsuperscript{3
\hyperref[_ENREF_23]{23}}) r\textsubscript{B} is \textasciitilde0.76 nm,
smaller than most proteins.

Figure 2A shows r\textsubscript{B} as a function of $\rho$\textsubscript{p}
at 298K. The data for the plot were calculated accounting for the
variation of $\eta$ (through
Andrade\textsuperscript{\hyperref[_ENREF_24]{24},\hyperref[_ENREF_25]{25}}
and Vogel\hyperref[_ENREF_26]{\textsuperscript{26}} equations using
parameters from Reid, Prausnitz \&
Poling\hyperref[_ENREF_27]{\textsuperscript{27}}) and $\rho$\textsubscript{s}
(through Hankinson-Brobst-Thomson
technique\hyperref[_ENREF_28]{\textsuperscript{28}} using parameters
also from Reid, Prausnitz \&
Poling\hyperref[_ENREF_27]{\textsuperscript{27}}) with T. 

\begin{figure}
    \centering
    \includegraphics[width=0.45\linewidth]{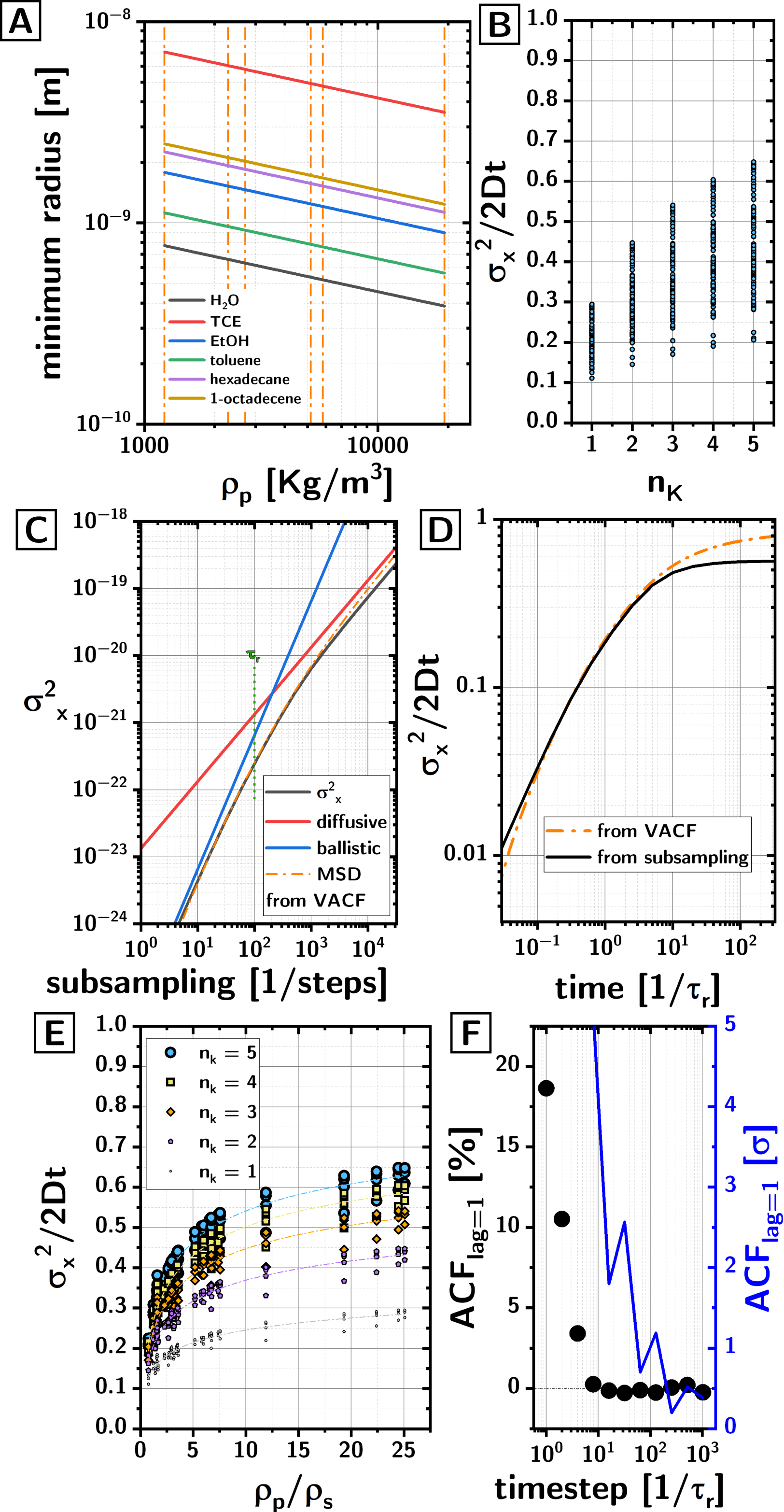}
    \caption{\textbf{(A)} Minimum radius of particles describable
as Brownian (i.e., $\tau$\textsubscript{\!r} \textgreater{}
10$\tau$\textsubscript{\!c}), as a function of their density, their phase, and
the solvent. \textbf{(B)} Ratio between the observed variance of the
step length $\sigma$\textsubscript{\!x}\textsuperscript{\!2} (single cartesian
coordinate) in the timelapse and the diffusional prediction 2Dt. The
ratio is plotted as a function of the time resolution of the timelapse
(physical trajectory is sampled at $\tau$\textsubscript{\!K} =
n\textsubscript{k}·$\tau$\textsubscript{\!r} time intervals). \textbf{(C)}
Variance of the step length $\sigma$\textsubscript{\!x}\textsuperscript{\!2} from
our timelapse (black line), compared to the diffusive MSD (red),
ballistic MSD (blue), and the MSD obtained from the VACF of the physical
trajectory (orange dash dot). The data is plotted as a function of the
time interval. The MSD obtained from the VACF transitions from ballistic
to diffusive regimes around the relaxation time $\tau$\textsubscript{\!r}, but
the variance of the step length $\sigma$\textsubscript{\!x}\textsuperscript{\!2}
from the timelapse remains offset from the diffusional prediction at
long times. \textbf{(D)} Same data as panel C but plotted in terms of
the ratio between the MSD and the diffusional prediction. \textbf{(E)}
Ratio between the MSD and the diffusional prediction as a function of
the density ratio ($\rho$\textsubscript{p}/$\rho$\textsubscript{s}) and the
subsampling rate. \textbf{(F)} Decorrelation at lag=1 as a function of
the timestep in the timelapse (in units of $\tau$\textsubscript{\!r}),
expressed either as percentage of autocorrelation (black scatters) or as
standard deviations from the mean autocorrelation at large lags (blue).}
    \label{two}
\end{figure}

Brownian dynamics \emph{can}, in principle, be used to describe the
motion in liquids of nearly everything but small molecules.

\textbf{The ``physical'' trajectory.} Simulations are discrete by their
own very nature. The smallest ``discontinuities'' in the motion of the
particles in solution are the individual collisions with the solvent
molecules. This means that choosing \emph{$\tau$\textsubscript{\!c}} as the
discrete time step used in calculating this ``physical'' trajectory
seems reasonable (cf. Figure 1A). Once determined the timescale, all one
needs to calculate the trajectory, in theory, is a suitable VACF.

\emph{The distribution of the displacement magnitudes of the physical
trajectory.}

The mean squared displacement
\emph{$\sigma$\textsubscript{x}}\textsuperscript{2} as a function of time can
be obtained directly from the VACF by

\[
\sigma_{x}^{2}(t) = 2 \int_{0}^{t} (t - \tau) C_{v}(\tau) \, d\tau \tag{6}
\]

where \emph{C\textsubscript{v}} is the non-normalized
VACF\hyperref[_ENREF_29]{\textsuperscript{29}}.

The VACF decays exponentially only if caused by random noise. In
reality, the interaction with the surrounding liquid during motion is
not uncorrelated\hyperref[_ENREF_30]{\textsuperscript{30}}. Firstly, as
the particle moves into the liquid, it ``writes'' information about its
own passage into the dynamics of the liquid itself: vortices (i.e., a
collective, correlated motion of the liquid molecules) are created in
the wake of the particle. A particle moving inside this correlated
velocity field is inevitably correlated with its own past motion. The
correlation vanishes in time as these vortices propagate away from the
particle over the ``vorticity diffusion time'',
\(\text{$\tau$}_{\text{v}}\text{=}\frac{\text{$\rho$}_{\text{s}}}{\text{$\eta$}}\text{r}_{\text{p}}^{\text{2}}\),
where $\rho$\emph{\textsubscript{s}} is the density of the liquid.

The second effect is due to the compressibility of the liquid, which
causes it to undergo compression when the particle pushes against it.
This compression establishes acoustic waves in the liquid that affect
the velocity field of the liquid surrounding the particle at later
times. The timescale of this process is the sonic time
\(\text{$\tau$}_{\text{s}}\text{=}\frac{\text{r}_{\text{p}}}{\text{c}}\),
where \emph{c} is the speed of sound in the liquid.

These two effects are usually somewhat separated in timescale (typically
\emph{$\tau$\textsubscript{s}} \textless{} \emph{$\tau$\textsubscript{v}} , cf.
Figure S3A), so they are accounted for by a VACF composed of two terms
that dominate respectively at long and short
timescales\hyperref[_ENREF_30]{\textsuperscript{30-34}}:

\begin{equation}
C_{v}(t) = \frac{k_{B} T}{M} \left\{ \frac{2 \rho_{p}}{9 \pi \rho_{s}} \int_{0}^{\infty} \frac{\sqrt{x} \, e^{-\frac{xt}{\tau_{v}}}}{1 + \sigma_{1} + \sigma_{2} x^{2}} \, dx + \frac{e^{\frac{\alpha_{1} t}{\tau_{s}}}}{1 + \frac{2 \rho_{p}}{\rho_{s}}} \left[ \cos\left(\frac{\alpha_{2} t}{\tau_{s}}\right) - \frac{\alpha_{1}}{\alpha_{2}} \sin\left(\frac{\alpha_{2} t}{\tau_{s}}\right) \right] \right\}
\tag{6}
\end{equation}

where
\(\text{$\sigma$}_{\text{1}}\text{=}\frac{\text{1}}{\text{9}}\left( \text{7-4}\frac{\text{$\rho$}_{\text{p}}}{\text{$\rho$}_{\text{s}}} \right)\),
\(\text{$\sigma$}_{\text{2}}\text{=}\frac{\text{1}}{\text{81}}\left( \text{1+2}\frac{\text{$\rho$}_{\text{p}}}{\text{$\rho$}_{\text{s}}} \right)^{\text{2}}\),
\(\text{$\alpha$}_{\text{1}}\text{=1+}\frac{\text{$\rho$}_{\text{s}}}{\text{2}\text{$\rho$}_{\text{p}}}\),
\(\text{$\alpha$}_{\text{2}}\text{=}\sqrt{\text{1-}\left( \frac{\text{$\rho$}_{\text{s}}}{\text{2}\text{$\rho$}_{\text{p}}} \right)^{\text{2}}}\),
and \emph{M} is the ``effective mass'' of the particle
\(\text{M=}\frac{\text{4}}{\text{3}}\text{$\pi$}\text{r}_{\text{p}}^{\text{3}}\left( \text{$\rho$}_{\text{p}}\text{+}\frac{\text{$\rho$}_{\text{s}}}{\text{2}} \right)\):
a particle moving a liquid must displace liquid to do so which means
that its mass must include the mass of the displaced liquid.

To generate the physical trajectory, I calculated this VACF for each
combination of solvent and particle up to times equal to
3·\emph{$\tau$\textsubscript{\!r}} in time intervals equal to
\emph{$\tau$\textsubscript{\!c}}. The MSD,
\emph{$\sigma$\textsubscript{x}}\textsuperscript{2}, at the time
\emph{$\tau$\textsubscript{\!c}} (i.e., the timestep chosen for the physical
trajectory) was calculated from the same VACF, but integrated with a
much finer time resolution
(\emph{$\tau$\textsubscript{\!c}}/10\textsuperscript{4}) given the sensitivity
of the integral in Equation 4 to the discretization (cf. Figure 1A). The
calculated VACF follows the expected power law tail with a -3/2 exponent
(Figure S3B).

\emph{The persistence of motion.}

A more delicate process is the determination of the displacements for
each timestep, since these have to be autocorrelated according to the
\emph{C\textsubscript{v}}(\emph{t}) I just described. To do so, it is
necessary to determine the autoregression coefficients
\emph{$\beta$\textsubscript{i}} for the expansion

\begin{equation}
v_{x}(t) = \sum_{i=1}^{l} \beta_{i} v(t-i) + \xi(t) \tag{7}
\end{equation}

where \emph{v\textsubscript{x}}(\emph{t}) is the \emph{x} component of
the velocity at the step \emph{t}, \emph{$\xi$}(\emph{t}) is the noise
function, and \emph{l} is the maximum lag being considered (in our case
3\emph{$\tau$\textsubscript{\!r}}/\emph{$\tau$\textsubscript{\!c}}). The coefficients
were calculated by solving the corresponding Yule-Walker system of
equations expressed in terms of the autocorrelation function

\begin{equation}
\begin{pmatrix}
C'_{1} \\
C'_{2} \\
C'_{3} \\
\vdots \\
C'_{l}
\end{pmatrix}
=
\begin{bmatrix}
C'_{0} & C'_{1} & C'_{2} & \ldots \\
C'_{1} & C'_{0} & C'_{1} & \ldots \\
C'_{2} & C'_{1} & C'_{0} & \ldots \\
\vdots & \vdots & \vdots & \ddots \\
C'_{l-1} & C'_{l-2} & C'_{l-3} & \ldots 
\end{bmatrix}
\begin{pmatrix}
\beta_{1} \\
\beta_{2} \\
\beta_{3} \\
\vdots \\
\beta_{l}
\end{pmatrix}
\end{equation}

where the left side is the vector of the normalized autocorrelation
function \emph{C'} (subscripts indicating the lag as number of time
intervals $\tau$\textsubscript{\!c}), while the right side is the product of
the Toeplitz matrix of \emph{C'} (from lag 0) and the vector of the
autoregression
coefficients\textsuperscript{\hyperref[_ENREF_35]{35},\hyperref[_ENREF_36]{36}}.

This brute force approach is not without challenges: for the
coefficients obtained from \emph{C'} to generate a time series with
autocorrelation close to \emph{C'} it is necessary for the input
\emph{C'} to be highly precise (i.e., integration of the VACF must be
done with small \emph{dx} intervals). Since many of our simulations
involve lags in excess of 10\textsuperscript{5} timesteps, solving the
system of equations in double precision can be cumbersome (since I am
modeling one particle at a time, the matrices are large but I could not
make GPU processing in double precision advantageous). Estimation
approaches such as the Durbin-Levinson did not give me coefficients able
to reproduce the desired autocorrelation in the produced time
series\hyperref[_ENREF_37]{\textsuperscript{37}}. Using this approach,
the VACF of the physical trajectory produced using the autoregression
coefficients as described above match closely the input VACF (cf. Figure
S3C).

Once I determined the \emph{$\beta$\textsubscript{i}} coefficients, the
individual components of the displacements were generated as a
timeseries of unitary variance by scaling the noise term
\emph{$\xi$}(\emph{t}) by a factor
\(\text{$\sigma$}_{\text{n}}\text{=}\sqrt{\text{1-C'}\text{$\cdot$}\text{$\beta$}}\), where
the dot indicates a dot product of the \emph{C'} and \emph{$\beta$} vectors.
The resulting time series of the individual components of the 3D
displacement shows the expected variance and distribution (cf. Figure
S3D). The time series is then scaled to obtain the variance consistent
with \emph{$\sigma$\textsubscript{x}}\textsuperscript{2} (cf. Figure 1A).

This process is executed for each spatial component, obtaining
independent timeseries for the individual components of the displacement
(cf. Figure 1A-B), which are then assembled as displacement vectors, and
their cumulative sum is the final trajectory.

The distribution of the lengths of the 3D displacement vectors follows
expectedly a Maxwell-Boltzmann distribution (cf. Figure S3E). The RMS
velocities obtained from these distributions are close to the ballistic
values as expected from the fact that the timestep of the physical
trajectory is $\tau$\textsubscript{\!c} (cf. Figure S3F). The difference from
the ballistic values depends on the relative solvent/particle
compositions and especially on their density ratio as it determines the
decorrelation impact associated with each solvent/particle collision.
Lastly, the orientations of the displacements that make up the physical
trajectory, sampled over a long enough time, appear to be ergodic (cf.
Figure S3G).

The autocorrelation functions at lag=1 associated with the physical
trajectory are not identical (cf. Figure S3H for an example). The
autocorrelation of the moduli of the displacement vectors is lower than
that for the individual components, while the autocorrelation of the
orientation of the displacement vectors is also lower than that for the
cartesian components of the displacements but higher than for the
moduli. In other words, the orientation of motion is more strongly
autocorrelated at short times than the lengths of the steps.

\textbf{Subsampling into a random flight.} While calculating the
physical trajectory (cf. Figure 1B for an example of a small section of
a calculated trajectory), I took ``snapshots'' of the particle positions
at intervals of time \emph{$\tau$\textsubscript{\!K}} where
\emph{$\tau$\textsubscript{\!K}} = \emph{n\textsubscript{K}}·\emph{$\tau$\textsubscript{\!r}} , where \emph{n\textsubscript{K}} is an
integer I call the ``Kuhn multiplier'' (in our simulations
\emph{$\tau$\textsubscript{\!K}} was between 1 and 5). The name ``Kuhn
multiplier'' is an homage to Prof. Werner Kuhn who applied a reasoning
similar to this to describe the conformation of polymers as random
flights with the goal of calculating their conformational
entropy\hyperref[_ENREF_38]{\textsuperscript{38}}.

As these snapshots separated by time intervals \emph{$\tau$\textsubscript{\!K}}
are compiled, the positions of the particle in those snapshots form what
looks like a random flight that is a coarser-grained (i.e., subsampled),
but accurate representation of the physical trajectory (cf. Figure 1).

In a sense this is the opposite conceptual approach that Prof. Kuhn
took. I am not creating random flights of fixed step length and try to
find at which value of step length the overall conformation of the
polymer is described by the random flight. I am taking the physical
trajectory, snapshot it at fixed time intervals, and find out what kind
of random flights are generated as a result.

For such extracted random flights to be then separately reproduced
\emph{ex novo} as a coarse-grained model of a physical Brownian
trajectory, they should be then characterized for: (i) their step length
distribution; (ii) their autocorrelation and autoregression
coefficients. In what follows I describe how these functions change as a
function of solvent, material and size of the particle, as well as
\emph{n\textsubscript{K}}.

\subsection{Results and Discussion}\label{results-and-discussion}

It could be tempting to think that the MSD, VACF and other quantities
associated with the timelapse would be just the values of MSD and VACF
from the physical trajectory at longer times. This is not the case.
Subsampling changes the autocorrelation of a timeseries in often
unpredictable ways, even if the overall conformation of the trajectory
is preserved\hyperref[_ENREF_39]{\textsuperscript{39}}. This can be
understood in terms of Fourier transforms: a subsampling is a low-pass
filter that ``forgets'' high frequency correlations in a way that
depends subtly on the structure of the VACF (which is why I could not
just calculate the MSD at t = $\tau$\textsubscript{\!K}).

The timelapses are essentially somewhat persistent random flights where
the steps are equispaced in time by a multiple of the relaxation time of
the system. To characterize these flights we have to look at the
distribution and mean squared values of the displacements, and the ACFs
of the displacements.

The distributions of the single component displacements in the timelapse
were exceedingly well fitted with normal distributions
(R\textsuperscript{2} across all our conditions was 0.9993±0.0001), and
the displacement step acceptably well described by Maxwell-Boltzmann
distributions (R\textsuperscript{2}=0.998±0.01).

There are two noteworthy observations.

The first is that the $\sigma$\textsubscript{\!x}\textsuperscript{\!2} of the
subsampled steps are distinctly subdiffusive in absolute value (the
exponent of the dependence on time is instead diffusive) even when the
snapshots are taken every 5$\tau$\textsubscript{\!r} (cf. Figure 2B, the single
step $\sigma$\textsubscript{\!x}\textsuperscript{\!2} (averaged across all our
simulations) was 0.44 times the diffusive value when
n\textsubscript{k}=5, i.e., when motion is often regarded as diffusive.
This is not necessarily surprising for at least two reasons: (i) at very
short timescales even ballistic motion is, mathematically speaking,
``subdiffusive'' and (ii) $\sigma$\textsubscript{\!x}\textsuperscript{\!2} of a
subsampled time series are basically moving averages with a window as
wide as the time interval, hence they ``lag'' the continuous time series
they derive from.

It is easy to show how, for an identical trajectory, the
$\sigma$\textsubscript{\!x}\textsuperscript{\!2} for a single step depends on its
subsampling. A single physical trajectory was calculated for
10\textsuperscript{9} steps corresponding to $\tau$\textsubscript{\!c} and
where $\tau$\textsubscript{\!r} $\cong$ 100$\tau$\textsubscript{\!c}. The trajectory was
then subsampled every 2, 4, 8 steps and so on, up to 32768 steps. The
variance of the Gaussian distribution of the single components of the
displacements (i.e., the $\sigma$\textsubscript{\!x}\textsuperscript{\!2}) in these
subsamplings can be then compared with the diffusive and ballistic MSD
obtaining the plot in Figure 2C. Importantly, the dash-dot orange curve
in Figure 2C shows the MSD one could have calculated directly from the
VACF at the corresponding time intervals. The discrepancy between the
MSD from the subsampling and that from the VACF becomes very significant
for subsamplings larger than $\tau$\textsubscript{r.}

Furthermore, the ratio between the variance
$\sigma$\textsubscript{\!x}\textsuperscript{\!2} of the distribution of single step
displacements of the subsampled trajectory and 2Dt is different from
what one would obtain directly from the VACF in Eq. 5 (cf. Figure S2D).
The $\sigma$\textsubscript{\!x}\textsuperscript{\!2} derived directly from the VACF
tends asymptotically to 2Dt as t increases, while the one obtained from
subsampled trajectories plateaus to a value that is significantly lower
than 2Dt.

This means that picking a time interval for a Brownian simulation that
is orders of magnitude larger than $\tau$\textsubscript{\!r} does not lead to a
diffusive step length as it is usually assumed on the basis of the
relationship between VACF and MSD. Making that approximation causes a
very significant overestimate of the step lengths of the random walk.

The second noteworthy thing is the large spread in the values of
$\sigma$\textsubscript{\!x}\textsuperscript{\!2}/2Dt for identical values of
n\textsubscript{k}. This could be surprising or not depending on the
reason why it happens. It could be because of (i) insufficient
statistics on the simulations, (ii) excessively coarse integration of
the VACF or premature truncation of it, or (iii) something intrinsic to
the model.

The first two hypotheses were easily disproven. First,
$\sigma$\textsubscript{\!x}\textsuperscript{\!2} calculated from 1000 replicate
trajectories gave relative standard errors on the mean of
10\textsuperscript{-4}: the values are reproducible. Second, extending
the calculation of the VACF to 100$\tau$\textsubscript{\!r} and decreasing the
size of the dx interval in Eq. 5 by three orders of magnitude did not
change the $\sigma$\textsubscript{\!x}\textsuperscript{\!2} significantly
(certainly less than 1\%): it is not a problem with the numerical
integration.

The explanation for the spread is hinted at by the fact that the
$\sigma$\textsubscript{\!x}\textsuperscript{\!2}/2Dt ratio appears to be dependent
on the density ratio $\rho$\textsubscript{p}/$\rho$\textsubscript{s} (cf. Figure
2E), without explaining fully the spread. The density ratio is featured
prominently in Eq. 5 which quantifies the VACF. The VACF in Eq. 5, as
already stated, does not decay exponentially. Nonetheless, the
$\tau$\textsubscript{\!r} on which I base the subsampling intervals is obtained
from Eq. 4 which assumes an exponential decay of the VACF. Furthermore,
the subsampling intervals I picked in order to allow for the study of
small Brownian particles, are in a range where the
$\sigma$\textsubscript{\!x}\textsuperscript{\!2}/2Dt ratio is still rapidly
changing (cf. Figure 2D). Taken together, these facts indicate that the
spread of the values in Figure 2B are the result of (i) the difficulty
of quantifying a relaxation time for an algebraic decay curve (i.e.,
$\tau$\textsubscript{\!r} corresponds to different degrees of decorrelation
depending on the density ratio and other physical variables used in Eq.
5), (ii) the steep dependence of the
$\sigma$\textsubscript{\!x}\textsuperscript{\!2}/2Dt ratio on time for subsampling
intervals of small multiples of $\tau$\textsubscript{\!r}.

An unfortunate consequence of this spread is that the VACF for the
subsampled random flight must be probably calculated before each
simulation.

Lastly but importantly: using the correct
$\sigma$\textsubscript{\!x}\textsuperscript{\!2} for the timestep chosen can be
sufficient in generating accurate \emph{ex novo} trajectories. To assess
how quickly the autocorrelation vanishes as a function of the
subsampling we calculated the autocorrelation of the 1D components of
the displacements as a function of the subsampling in units of
$\tau$\textsubscript{\!r} (for the calculation we used the
10\textsuperscript{9} step physical trajectory mentioned above). This
analysis (cf. Figure 2F) shows that autocorrelation at lag=1 vanishes
for subsamplings where the timestep is approximately equal to
32$\tau$\textsubscript{\!r}. If one looks at how the autocorrelation at lag=1
compares to the standard deviation of the autocorrelation at large lags
(cf. Figure 2F, blue) then one might want to push to
200$\tau$\textsubscript{\!r} to ensure complete loss of autocorrelation.

Somewhat surprisingly, in most cases for which $\tau$\textsubscript{\!r}
\textgreater{} 10$\tau$\textsubscript{\!c} (i.e., Brownian regime) the
condition for memorylessness of the path (timestep = 200
$\tau$\textsubscript{\!r}) leads to RMS step lengths safely shorter than the
diameter of the particles (cf. Figure 3A-E). In other words, a Brownian
simulation where the timestep is commensurate to the time it takes the
particle to move by its radius can safely ignore autocorrelation. It is
obviously important to keep the Brownian step significantly smaller than
the particle radius since the steps are normally distributed and
Brownian simulations involve often large numbers of steps: steps that
are 5+ standard deviations away from the mean cannot be excluded. This
fact, especially in the case of highly concentrated systems, can cause
issues in the accurate detection of collisions.

In the case in which instead the timestep chosen is a small multiple of
$\tau$\textsubscript{\!r}, then autocorrelation must be considered. To do so,
we took the subsampled trajectories, extracted their VACF, calculated
the autoregression coefficients by solving Eq. 7, and then used Eq. 6 to
calculate ex novo new trajectories. The noise term was normalized to
give a variance equal to the $\sigma$\textsubscript{\!x}\textsuperscript{\!2}
extracted from the original subsampled trajectory.

The results of this approach are positive. The VACF obtained from these
ex-novo trajectories match very closely the VACF of the original
subsampled trajectory (cf. Figure 3F for a representative example).

\begin{figure}
    \centering
    \includegraphics[width=1\linewidth]{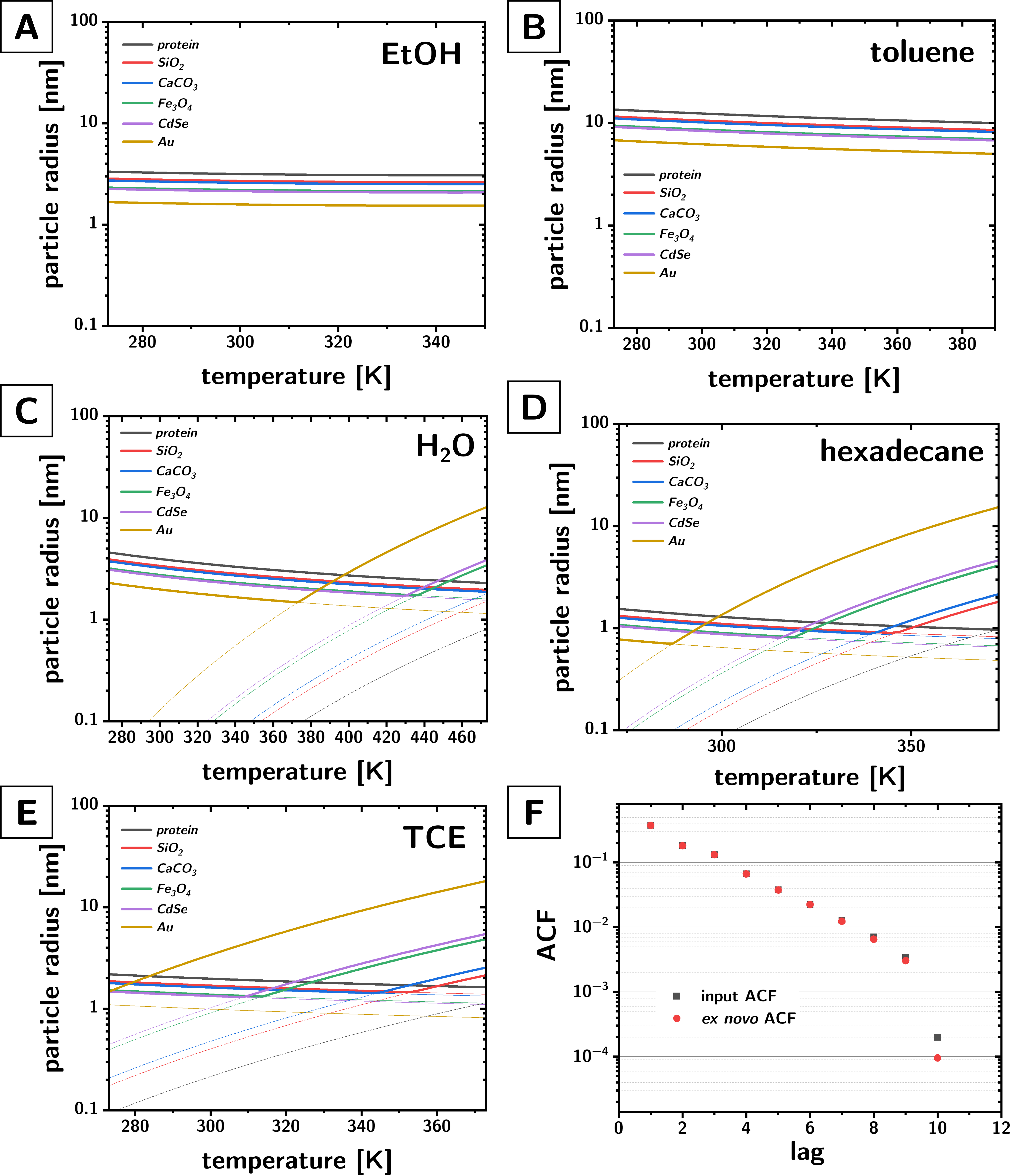}
    \caption{\textbf{(A-E)} Minimum particle radii (as a function
of particle phase, temperature, and solvent) to satisfy the Brownian
regime (thin colored solid lines) and the memorylessness approximation
(dash-dotted thin lines), i.e., timesteps long enough to eliminate
autocorrelation ($\tau$\textsubscript{\!K}\textasciitilde{}
200$\tau$\textsubscript{\!r}) produce RMS displacements significantly shorter
than the diameter of the particles. \textbf{(F)} Representative
comparison between the autocorrelation (ACF) obtained from the input
trajectory (i.e., the timelapse of the physical trajectory), and from
the ex-novo trajectories obtained with the autoregression coefficients
extracted from the input ACF.}
    \label{three}
\end{figure}

\subsection{Conclusions}\label{conclusions}

In conclusion, I have explored and discussed to what extent a Brownian
trajectory can be modeled as a random flight. The reason why I did it is
that diffusion is often modeled (especially in simulations of ensembles
of particles) as a random flight assuming memoryless trajectories and
diffusional step lengths ($\sigma$\textsubscript{x} =
(2Dt)\textsuperscript{1/2}) on top of which corrections are then made to
account for interactions. As I show here, the first approximation can be
often reasonable, but the second is fundamentally wrong and leads to a
very significant overestimation of the step lengths,

To find this, I have (i) calculated a physical Brownian trajectory by
considering collisional timescales and a VACF that accounts for
hydrodynamic and acoustic effects, (ii) taken snapshots of particle
positions at time intervals that are multiples of the relaxation time,
(iii) analyzed the subsampled trajectory as an autocorrelated random
flight extracting MSD and VACF.

The simulations (considering a variety of solvents, particle phases, and
particle sizes) show that (i) subsampled flights become genuinely
memoryless only when the time step is \textasciitilde200 times larger
than the relaxation time, and (ii) the distribution of the step lengths
is nicely gaussian (on each cartesian component) but with variances that
remain significantly smaller than 2Dt. As we discuss, this last fact is
due to the combination of two effects: diffusional MSD is mathematically
superballistic at short timescales and subsampled trajectories are
moving averages of the underlying physical trajectory. In other words,
maybe counterintuitively, the MSD of the physical trajectory at long
time intervals asymptotically approaches 2Dt, but the MSD of the
subsampled steps does not, even when the associated time interval is
several hundred times larger than the relaxation time.

Practically speaking, given the inaccuracies of the estimated relaxation
times, the appropriate corrections to the step lengths (or,
alternatively, to the diffusivities) must be extracted from a simulated
physical trajectory, using the physical parameters of the system at hand
and the chosen subsampling. The simulations discussed here in fact show
that the precise value of the correction depends intimately on the
specific physical parameters (densities, radii, etc\ldots) in ways that
are not accurately captured by empirical fits.

\clearpage 

\section{Supporting Information}\label{supporting-information}

\clearpage 

\begin{figure}[h] 
    \centering
    \includegraphics[width=0.8\linewidth]{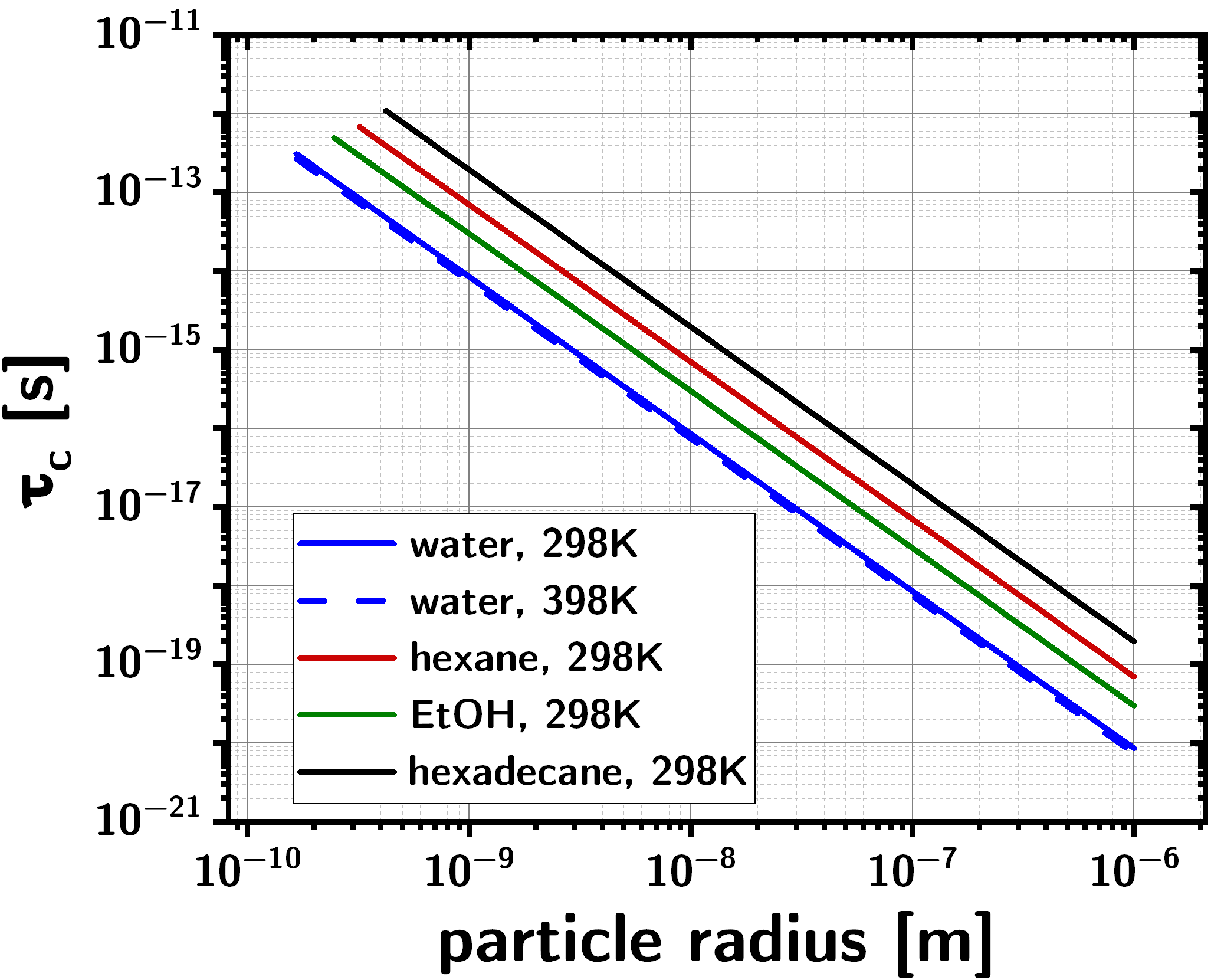}
    
    \begin{flushleft}
    \textbf{Figure S1. \emph{Collisional timescale as a function of solvent/particle characteristics and temperature.}}
    The calculated value of the collisional timescale $\tau$\textsubscript{\!c} (according to Eq. 3) as a function of the particle radius, for a range of solvents, temperatures, and particle phases.
    \end{flushleft}

    \label{fig:S1} 
\end{figure}

\clearpage 
\begin{figure}
    \centering
    \includegraphics[width=1\linewidth]{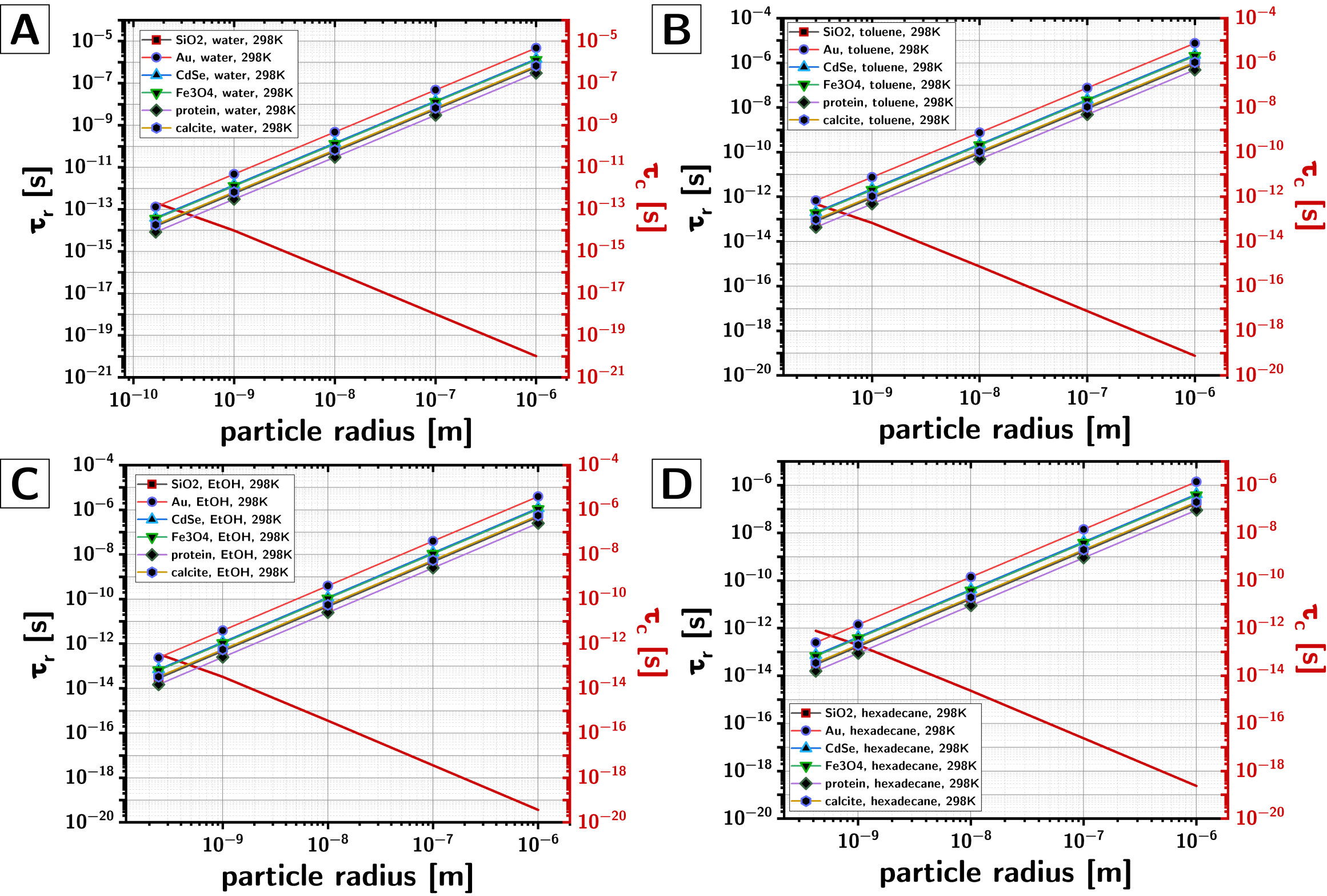}
    \begin{flushleft}
    \textbf{Figure S2. \emph{Collisional and relaxation timescales as a
function of particle radii, particle phases and solvents.}} The graphs
overlay, for the same system, the collisional timescale from Eq. 3 (red
line, no scatter, scale on the left axis) with the relaxation timescale
from Eq. 4 (colored lines, black-filled scatters, scale on the right
axis). As discussed in the main text, the validity of the relaxation
time formalism (i.e., $\tau$\textsubscript{\!r}$\gg$$\tau$\textsubscript{\!c}) falls
apart only when particles are smaller than
\textasciitilde10\textsuperscript{0} nm, depending on the combination of
solvent, particle phase. Data for different solvents are in different
panels( water (A), toluene (B), EtOH (C), and hexadecane (D)).
    \end{flushleft}
    
    \label{fig:S2}
\end{figure}

\clearpage 
\begin{figure}
    \centering
    \includegraphics[width=0.75\linewidth]{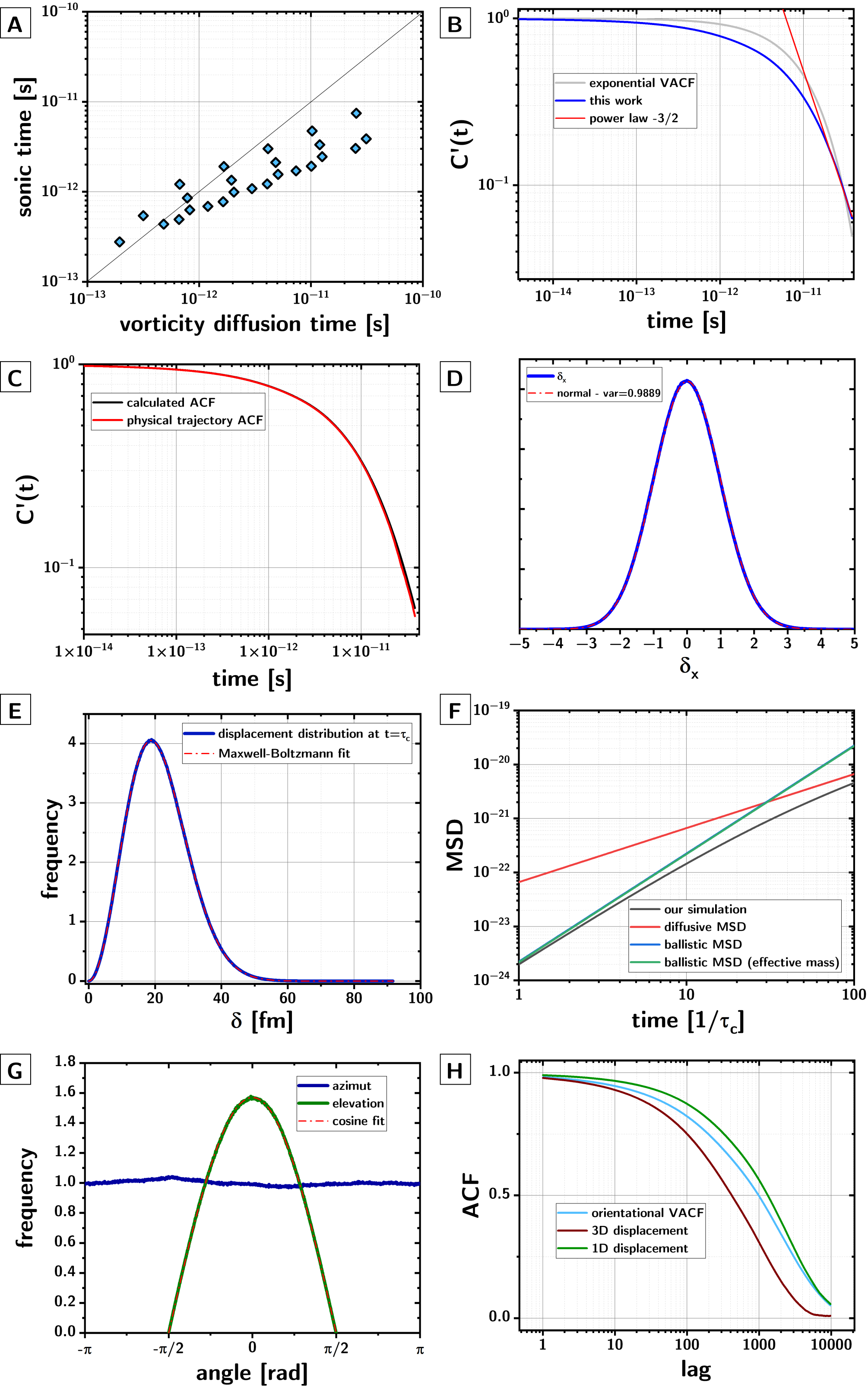}
    \begin{flushleft}
    \textbf{Figure S3. \emph{Computational diagnostics. A}.} Relation
between sonic time and vorticity diffusion time for the simulation
parameters explored. \textbf{B.} Comparison between normalized VACFs
(exponential in grey vs according to Eq. 5 in blue); the red line shows
the algebraic decay fit with a -3/2 exponent. \textbf{C.} Comparison
between the VACF calculated from Eq. 5 (black) and the one derived from
the physical trajectory generated from the $\beta$\textsubscript{i}
coefficients (red). \textbf{D.} Distribution of the time series of
one-dimensional univariant displacements in the physical trajectory
(blue) and the gaussian fit (red dashed). \textbf{E.} Distribution of
the 3D displacements magnitudes in a physical trajectory (blue) and
Maxwell-Boltzmann fit (red dashed). \textbf{F.} Comparison between the
MSD as a function of time (in $\tau$\textsubscript{\!c} units) from a physical
trajectory derived from the VACF in Eq. 5 (black), from a ballistic
motion (blue), ballistic motion accounting for the effective mass
(green) and from a diffusive motion (red). \textbf{G.} Distributions of
azimuths and elevations in the displacement orientations in a physical
trajectory. The dependence on the angle of both is consistent with a
uniform distribution on the spherical angle. \textbf{H.} Comparison of
the ACF (at lag=1) calculated from a physical trajectory (orientational
in cyan, 3D displacement magnitude in dark red, and 1D displacement
magnitude in green)
    \end{flushleft}
    \label{fig:S3}
\end{figure}

\clearpage 
\section{References}\label{references}

\phantomsection\label{_ENREF_1}{} (1) Gupta, C.; Sarkar, D.; Tieleman,
D. P.; Singharoy, A. \emph{Curr. Opin. Struct. Biol.} \textbf{2022},
\emph{73}, 102338.

(2) Lbadaoui-Darvas, M.; Garberoglio, G.; Karadima, K. S.; Cordeiro, M.
N. D.; Nenes, A.; Takahama, S. \emph{Mol. Simul.} \textbf{2023},
\emph{49}, 1229.

(3) Joshi, S. Y.; Deshmukh, S. A. \emph{Mol. Simul.} \textbf{2021},
\emph{47}, 786.

\phantomsection\label{_ENREF_4}{} (4) Muñiz‐Chicharro, A.; Votapka, L.
W.; Amaro, R. E.; Wade, R. C. \emph{Wiley Interdisciplinary Reviews:
Computational Molecular Science} \textbf{2023}, \emph{13}, e1649.

\phantomsection\label{_ENREF_5}{} (5) Anderson, J. A.; Lorenz, C. D.;
Travesset, A. \emph{J. Comput. Phys.} \textbf{2008}, \emph{227}, 5342.

\phantomsection\label{_ENREF_6}{} (6) Gkeka, P.; Stoltz, G.; Barati
Farimani, A.; Belkacemi, Z.; Ceriotti, M.; Chodera, J. D.; Dinner, A.
R.; Ferguson, A. L.; Maillet, J.-B.; Minoux, H. \emph{J. Chem. Theory
Comput.} \textbf{2020}, \emph{16}, 4757.

\phantomsection\label{_ENREF_7}{} (7) Smith, G. E.; Seth, R.
\emph{Brownian motion and molecular reality}; Oxford University Press,
2020.

\phantomsection\label{_ENREF_8}{} (8) Lucretius, T. C. \emph{De rerum
natura}, 58 BC.

\phantomsection\label{_ENREF_9}{} (9) Einstein, A. \emph{Ann. Phys}
\textbf{1906}, \emph{19}, 371.

(10) Smoluchowski, M. M. \emph{Bulletin International de
l\textquotesingle Académie des Sciences de Cracovie} \textbf{1906}, 202.

(11) Perrin, J. B. \emph{Ann. Chim. Phys.} \textbf{1909}, \emph{18}, 5.

\phantomsection\label{_ENREF_12}{} (12) Zaccone, A.; Dorsaz, N.; Piazza,
F.; De Michele, C.; Morbidelli, M.; Foffi, G. \emph{The Journal of
Physical Chemistry B} \textbf{2011}, \emph{115}, 7383.

(13) Pelargonio, S.; Zaccone, A. \emph{Physical Review E} \textbf{2023},
\emph{107}, 064102.

(14) Jain, R.; Sebastian, K. \emph{The Journal of chemical physics}
\textbf{2017}, \emph{146}.

(15) Lattuada, M.; Wu, H.; Morbidelli, M. \emph{Langmuir} \textbf{2004},
\emph{20}, 5630.

\phantomsection\label{_ENREF_16}{} (16) Selmeczi, D.; Tolić-Nørrelykke,
S.; Schäffer, E.; Hagedorn, P.; Mosler, S.; Berg-Sørensen, K.; Larsen,
N. B.; Flyvbjerg, H. \emph{Acta Physica Polonica B} \textbf{2007},
\emph{38}.

(17) Huang, R. \emph{Brownian motion at fast time scales and thermal
noise imaging}; The University of Texas at Austin, 2008.

(18) Li, T.; Raizen, M. G. \emph{Annalen der Physik} \textbf{2013},
\emph{525}, 281.

\phantomsection\label{_ENREF_19}{} (19) Philipse, A. \emph{Utrecht
University, Debye Institute, Van't Hoff Laboratory} \textbf{2011}.

\phantomsection\label{_ENREF_20}{} (20) Zaccone, A. \emph{Phys. Rev.
Lett.} \textbf{2022}, \emph{128}, 028002.

\phantomsection\label{_ENREF_21}{} (21) Chen, J. C.; Kim, A. S.
\emph{Adv. Colloid Interface Sci.} \textbf{2004}, \emph{112}, 159.

\phantomsection\label{_ENREF_22}{} (22) Huber, G. A.; McCammon, J. A.
\emph{Trends in chemistry} \textbf{2019}, \emph{1}, 727.

\phantomsection\label{_ENREF_23}{} (23) Andersson, K.; Hovmöller, S.
\emph{Zeitschrift für Kristallographie-Crystalline Materials}
\textbf{1998}, \emph{213}, 369.

\phantomsection\label{_ENREF_24}{} (24) Andrade, E. d. C. \emph{The
London, Edinburgh, and Dublin Philosophical Magazine and Journal of
Science} \textbf{1934}, \emph{17}, 497.

\phantomsection\label{_ENREF_25}{} (25) Andrade, E. d. C. \emph{Nature}
\textbf{1930}, \emph{125}, 309.

\phantomsection\label{_ENREF_26}{} (26) Vogel, H.; Weiss, A.
\emph{Berichte der Bunsengesellschaft für physikalische Chemie}
\textbf{1982}, \emph{86}, 193.

\phantomsection\label{_ENREF_27}{} (27) Reid, R.; Prausnitz, J. P., BE;
McGraw-Hill: 1987.

\phantomsection\label{_ENREF_28}{} (28) Hankinson, R. W.; Thomson, G. H.
\emph{AIChE J.} \textbf{1979}, \emph{25}, 653.

\phantomsection\label{_ENREF_29}{} (29) Berlinsky, A.; Harris, A.
\emph{Statistical Mechanics: An Introductory Graduate Course}; Springer
Nature, 2019.

\phantomsection\label{_ENREF_30}{} (30) Chakraborty, D. \emph{The
European Physical Journal B} \textbf{2011}, \emph{83}, 375.

(31) Zwanzig, R.; Bixon, M. \emph{Phys. Rev. A} \textbf{1970}, \emph{2},
2005.

(32) Chow, T.; Hermans, J. \emph{Physica} \textbf{1973}, \emph{65}, 156.

(33) Hinch, E. J. \emph{J. Fluid Mech.} \textbf{1975}, \emph{72}, 499.

(34) Clercx, H.; Schram, P. \emph{Phys. Rev. A} \textbf{1992},
\emph{46}, 1942.

\phantomsection\label{_ENREF_35}{} (35) Brockwell, P. J. \emph{Time
series: Theory and methods}; Springer-Verlag, 1991.

\phantomsection\label{_ENREF_36}{} (36) Shumway, R. H.; Stoffer, D. S.;
Stoffer, D. S. \emph{Time series analysis and its applications};
Springer, 2000; Vol. 3.

\phantomsection\label{_ENREF_37}{} (37) Cybenko, G. \emph{SIAM Journal
on Scientific and Statistical Computing} \textbf{1980}, \emph{1}, 303.

\phantomsection\label{_ENREF_38}{} (38) Kuhn, W.
\emph{Kolloid-Zeitschrift} \textbf{1934}, \emph{68}, 2.

\phantomsection\label{_ENREF_39}{} (39) Kennedy, M. C.; Prichard, S. J.
\emph{Landsc. Ecol.} \textbf{2017}, \emph{32}, 945.

\end{document}